\documentclass[twocolumn,prb,amssymb,showpacs,showkeys]{revtex4}
\usepackage{graphicx}
\usepackage{dcolumn}
\usepackage{bm}
\begin{document}
\title{Pseudogap behavior of phase-separated Sm$_{1-x}$Ca$_x$MnO$_3$ : A comparative photoemission study with double exchange}
\author{P. Pal, M. K. Dalai, R. Kundu and B. R. Sekhar}
\email{sekhar@iopb.res.in}
\affiliation{Institute of Physics, Sachivalaya Marg, Bhubaneswar 751 005, India.}
\author{C. Martin}
\affiliation{Laboratoire CRISMAT, UMR 6508, ISMRA, Boulevard du Marechal Juin, 
14050 Caen, France}
\begin{abstract}

Using valence band photoemission we have demonstrated the presence of a pseudogap 
in the near Fermi level electronic spectrum of some of the mixed phase 
compositions of Sm$_{1-x}$Ca$_x$MnO$_3$ system. The pseudogap was found to grow in 
size over a large region of the phase diagram of this system, finally leading to a 
metal-insulator transition. We have made a study comparing the near Fermi 
level behaviors of this system to those of a canonical double exchange system, namely, 
La$_{1-x}$Sr$_x$MnO$_3$. This study intends to highlight one of the important 
differences between the phase separated and double exchange colossal 
magnetoresistance systems in the nature of their energy gaps across the 
metal-insulator transitions. These differences could be ascribed to the distortions 
in the MnO$_6$ octahedra of their structures that regulate the localization of 
charge carriers. We have discussed our results from the point of view of models based 
on the idea of phase separation. 

\end{abstract}
\maketitle

\section{INTRODUCTION}

Although, the double-exchange (DE) mechanism \cite{zener,anderson,gennes} could 
qualitatively explain the phenomenon of colossal magneto-resistance (CMR) in 
manganites, it was found insufficient to provide a consolidated picture 
accommodating their intricate transport and magnetic properties. Most of the 
alternate theories proposed recently \cite{millis,tvr}, are based on lattice 
polarons of one type or the other highlighting the strong electron-lattice 
interactions in these systems. These polarons and thereby most of the properties 
of these Mn$^{3+}$ - Mn$^{4+}$ mixed valent compounds depend on the topology of 
the MnO$_6$ octahedra in their structure. Among the recently proposed models, the 
one based on electronic phase separation (PS)\cite{moreo1,moreo2} has been 
attracting considerable attention. According the PS model the metal - insulator 
(MI) transitions in these materials are driven by the percolation of current 
through ferromagnetic metallic (FMM) domains embedded in an antiferromagnetic 
insulating (AFMI) matrix. There have been many structural studies showing the 
co-existence of such metallic and insulating phases 
\cite{simon,martin1,ling,radaelli} with sizes varying from nanoscopic to 
microscopic. The cluster-glass (CG) compositions of Sm$_{1-x}$Ca$_x$MnO$_3$ system 
are some of the materials in which such a phase separation could be unambiguously 
shown using neutron diffraction, owing to the large sizes of the magnetic domains 
possible in them \cite{martin1}. Separation of these magnetic phases in this 
system has been found to be related to the strong distortions in their 
GdFeO$_3$-type structure. Also, these distortions control the topology of the 
MnO$_6$ octahedra of their structure and thereby the one electron band width 
(W) and the electron localization effects. The near Fermi level (E$_F$) electronic 
behavior of these phase separated systems could possibly be different from those 
of the La$_{1-x}$Sr$_x$MnO$_3$ which has long been identified as a canonical DE 
CMR system \cite{tokura} due to its large W. Further, since the term J$_H$ $/$W 
(J$_H$ is the Hund's coupling) which expresses the effective coupling between the 
$e_g$ and the t$_{2g}$ electrons of the crystal field split MnO$_6$ octahedra in 
their structure, remains in the weak-coupling regime, the La$_{1-x}$Sr$_x$MnO$_3$ 
system is regarded as the least affected by the electron-electron and 
electron-lattice correlation effects among the CMR manganites. On the other hand, 
the Sm$_{1-x}$Ca$_x$MnO$_3$ with its small W and distorted MnO$_6$ octahedra, is 
more prone to the electron correlation effects such as the localization of charge 
carriers.

Recently, there have been many photoemission studies \cite{ebata1,pal,ebata2} 
highlighting the subtle changes in the near E$_F$ density of states (DOS) 
associated with the MI transitions in phase separated CMR systems. These shifts in 
the near E$_F$ spectral weights lead to charge order (CO) gaps and pseudogaps. 
Such pseudogaps which have been shown to be a generic behavior of CMR systems with 
mixed phases\cite{moreo1,pal} are closely related to the changes in the W and 
electron localization. As mentioned before, one expects distinct differences 
between the mixed phase Sm$_{1-x}$Ca$_x$MnO$_3$ and the DE driven 
La$_{1-x}$Sr$_x$MnO$_3$ systems in the nature of their energy gaps across the MI 
transitions. Ultraviolet photoelectron spectroscopy (UPS) is a powerful technique 
capable of probing the fine changes in the valence band electronic structure. In 
this paper, we show some of the important differences between the 
Sm$_{1-x}$Ca$_x$MnO$_3$ and the La$_{1-x}$Sr$_x$MnO$_3$ systems in their near 
E$_F$ electronic structure using high resolution UPS. The doping dependent 
pseudogap behavior observed in Sm$_{1-x}$Ca$_x$MnO$_3$ system is discussed here 
and compared with that of the DE driven La$_{1-x}$Sr$_x$MnO$_3$.

\section{EXPERIMENTAL}

Polycrystalline samples of Sm$_{1-x}$Ca$_x$MnO$_3$ and La$_{1-x}$Sr$_x$MnO$_3$ 
systems were prepared by solid state reactions by mixing MnO$_2$, CaO and 
Sm$_2$O$_3$ or La$_2$O$_3$ in stoichiometric proportions. The powders were first 
heated at 1000 C with intermediate grindings and then pressed in the form of 
pellets. They were then sintered at 1500 C for 12 h in air with a slow cooling 
down to 800 C and finally quenched to room temperature. Details of the sample 
preparation technique could be found elsewhere \cite{jmm}. The monophasic, 
homogeneous nature of the samples have been checked by x-ray powder and electron 
diffraction techniques. The cationic compositions, close to their nominal values 
were confirmed using energy dispersive spectroscopy and iodometric titrations. 
Magnetic and electrical transport properties of the samples were determined using 
a magnetometer (equipped with a superconducting quantum interference device) and 
four probe resistivity measurements. A co-existence of two separate phases of 
FMM and AFMI domains below $100$ K in $x = 0.85$ and $0.9$ of the 
Sm$_{1-x}$Ca$_x$MnO$_3$ system were shown earlier using neutron diffraction 
studies carried out at LLB, Saclay, France on the G41 diffractometer 
\cite{martin1,jmm}. Consolidated results of these studies is published eleswhere 
\cite{martin1,jmm,martin2}.

\begin{figure}[t]
\vskip 1.0cm
\includegraphics[width=3.0in]{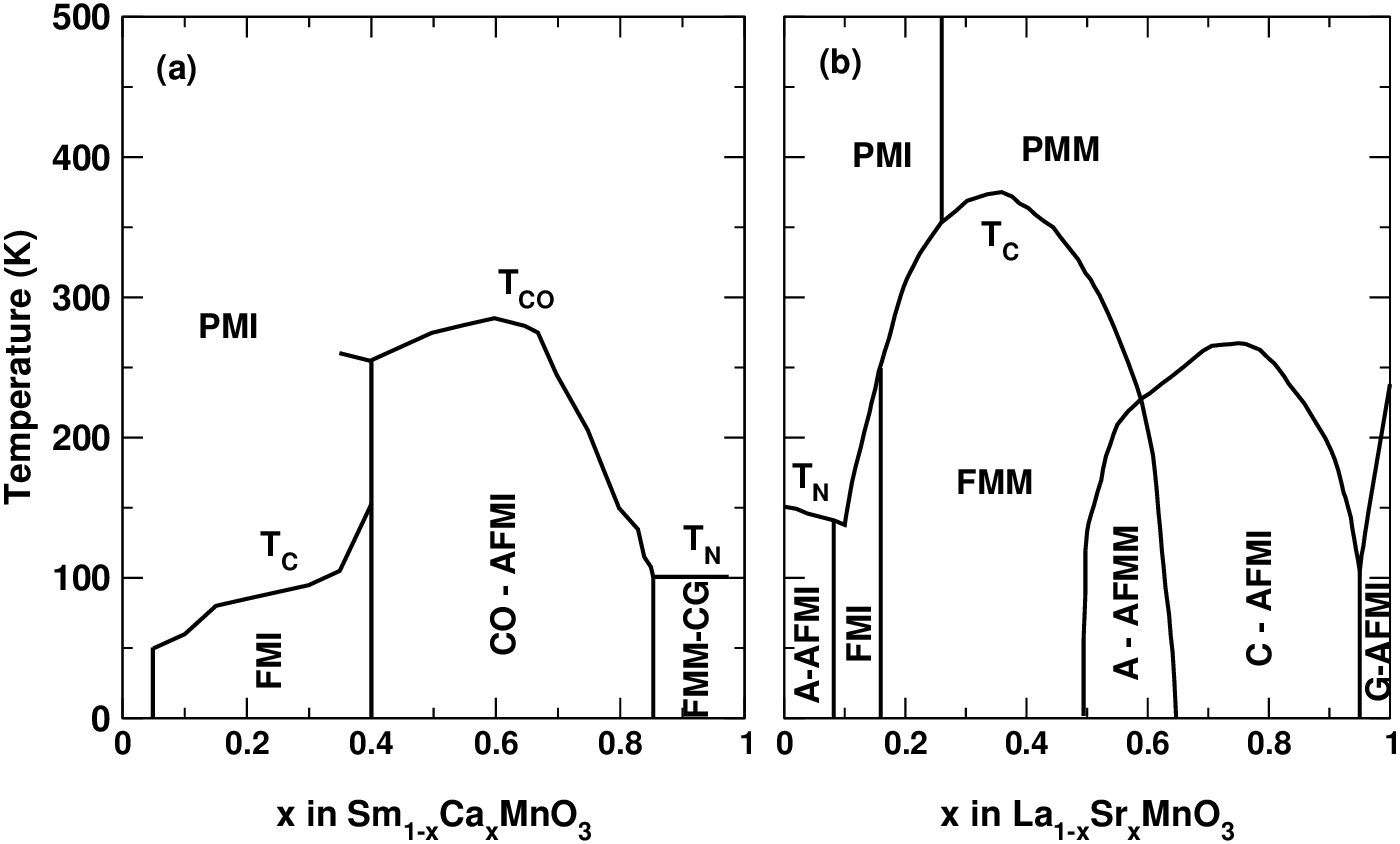}
\caption{\label{figure_1} Panel (a): Phase diagram of Sm$_{1-x}$Ca$_x$MnO$_3$ 
system. (b): Phase diagram of La$_{1-x}$Sr$_x$MnO$_3$ system adapted from 
the work of Chmaissem et al. \cite{chmaissem}.}
\end{figure}

Angle integrated ultraviolet photoemission measurements were performed using an 
Omicron mu-metal ultra high vacuum system equipped with a high intensity 
vacuum-ultraviolet source (HIS $13$) and a hemispherical electron energy analyzer 
(EA $125$ HR). At the He $I$ ($h$ $\nu$ = $21.2$ eV) line, the photon flux was of 
the order of $10^{16}$ photons/sec/steradian with a beam spot of $2.5$ mm 
diameter. Fermi energies for all measurements were calibrated using a freshly 
evaporated Ag film on a sample holder. The total energy resolution, estimated from 
the width of the Fermi edge, was about $80$ meV for He $I$ excitation. All the 
photoemission measurements were performed inside the analysis chamber under a base 
vacuum of $\sim$ $5.0$ $\times$ $10^{-11}$ mbar. The polycrystalline samples were 
repeatedly scraped using a diamond file inside the preparation chamber with a base 
vacuum of $\sim$ $5.0$ $\times$ $10^{-11}$ mbar and the spectra were taken within 
$1$ hour, so as to avoid any surface degradation. All measurements were repeated 
many times to ensure the reproducibility of the spectra. For the temperature 
dependent measurements, the samples were cooled by pumping liquid nitrogen through 
the sample manipulator fitted with a cryostat. Sample temperatures were measured 
using a silicon diode sensor touching the bottom of the stainless steel sample 
plate. The low temperature photoemission measuements at $77$ K were performed 
immediately after the cleaning the sample surfaces followed by the room 
tempedrature measurements. In order to make a good comparative study of the 
two systems (Sm$_{1-x}$Ca$_x$MnO$_3$ and La$_{1-x}$Sr$_x$MnO$_3$), we performed 
all the measurements under exactly same experimental conditions.

\begin{figure}[t]
\vskip 1.0cm
\includegraphics[width=3.0in]{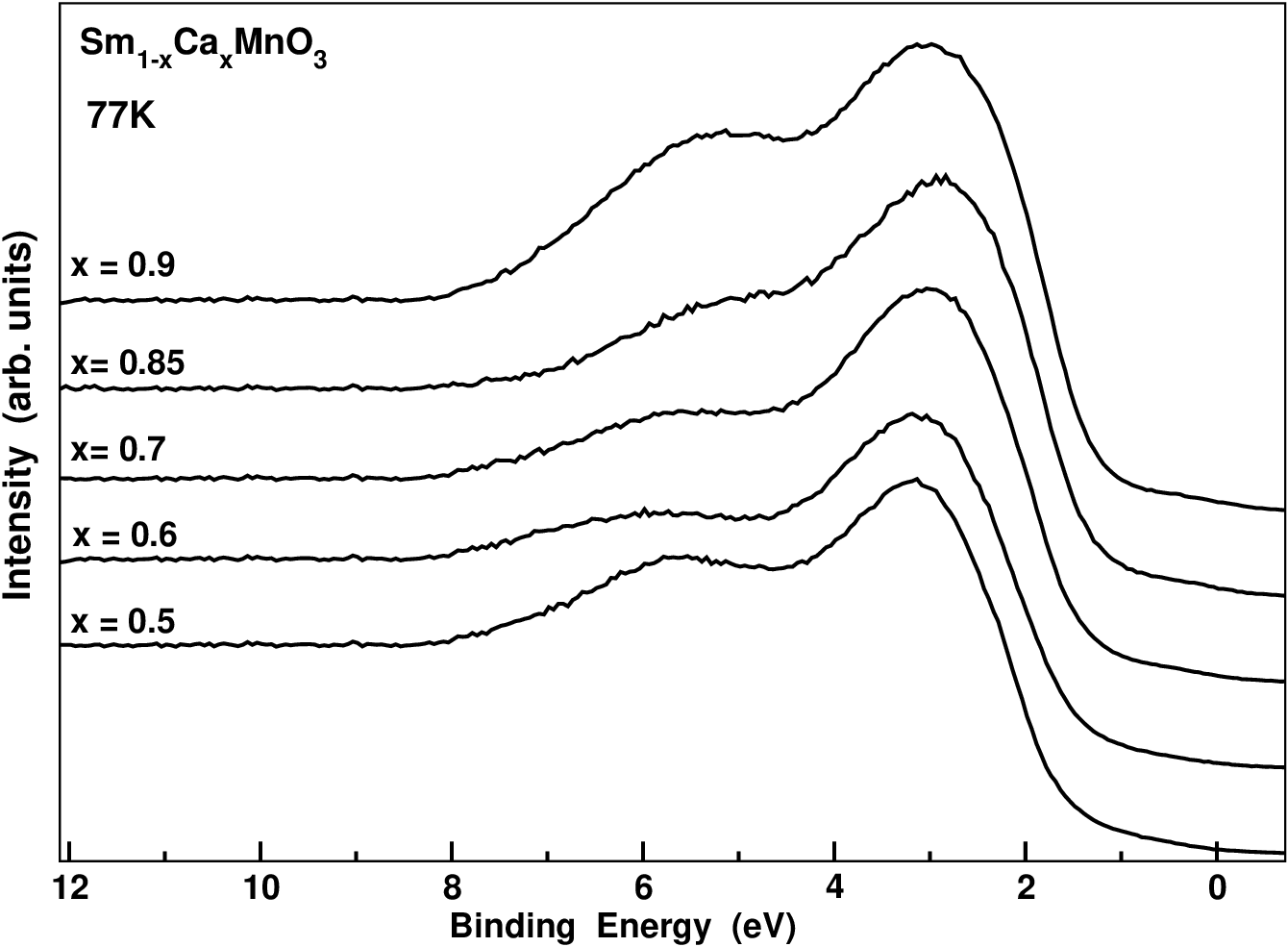}
\caption{\label{figure_2} The angle integrated valence band spectrum of the 
Sm$_{1-x}$Ca$_x$MnO$_3$ system taken at $77$ K using He I photon energy.}
\end{figure}

\section{RESULTS AND DISCUSSION}

The phase diagram of the Sm$_{1-x}$Ca$_x$MnO$_3$ system published earlier 
\cite{martin2} is shown in Fig. 1 (a). We have chosen five compositions of this 
system with $x$ = $0.3$, $0.5$, $0.7$, $0.85$ and $0.9$, which are all 
paramagnetic insulting (PMI) at room temperature. Below $100$ K, the $x = 0.3$
sample shows a ferromagnetic insulating (FMI) behavior. A large doping region, 
$0.40 \leqslant x \leqslant 0.7$, shows a CO-AFMI state at low temperatures. The 
CMR properties of these compositions are strongly dominated by this CO insulating 
state. The low temperature resistivities (down to $10$ K) of these samples 
were very high ($> 10^4$ $\Omega$ cm) and are thus insulators. The coexistence of 
ferromagnetism and metallicity is observed only in the cluster-glass (CG) region 
$i.e.$, $x \approx 0.9$, where the resistivity at low temperature (10 K) is 
$10^{-2}$ to $10^{-3}$ $\Omega$ cm, corresponding to that of 'bad' metals. The CMR 
effect is observed in the boundary between CG - state ($x>0.85$) and CO-AFMI 
state ($x<0.85$). Neutron diffraction studies\cite{martin1,jmm} have shown that 
the last two CG compositions ($x$ = $0.85$ and $0.9$) have coexisting FMM and AFMI 
phases. Fig. 1.(b) shows the phase diagram of the La$_{1-x}$Sr$_x$MnO$_3$ system, 
adapted from the work of Chmaissem et al.\cite{chmaissem}.

\begin{figure}[t]
\vskip 1.0cm
\includegraphics[width=3.0in]{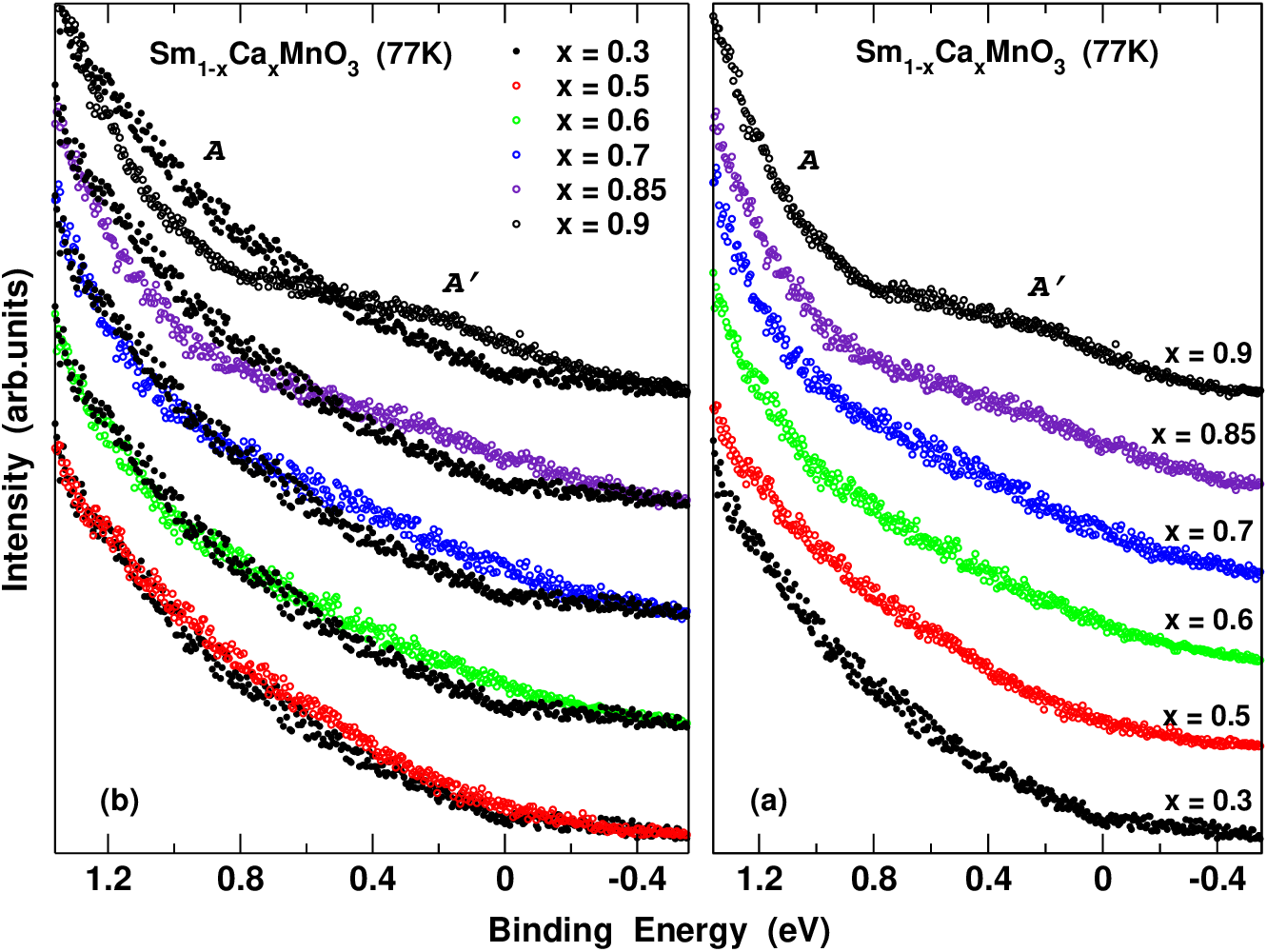}
\caption{\label{figure_3} Panel (a): High resolution photoemission spectra of the 
near E$_F$ region taken at $77$ K. A and A' refer to the Mn 3d e$_g$ spin 
up states of the system. The doping dependent MI transition is accompanied by the 
shifting of some DOS from A' to A, referred as the pseudogap in the text. (b): All 
the other spectra plotted against the one from $x$ = $0.3$ depicting the $x$ 
dependent growth of A'.} 
\end{figure}

In figure 2 we present the angle integrated valence band spectra of the 
Sm$_{1-x}$Ca$_x$MnO$_3$ samples taken at $77$ K. The main features seen in these 
spectra are by now well-known and looks similar to those reported earlier on 
different CMR systems \cite{saitoh,sarma,manas}, including the 
La$_{1-x}$Sr$_x$MnO$_3$. Features seen in the spectra originate from the bonding 
and antibonding states of Mn 3$d$ - O 2$p$ hybridization. Detailed discussion on 
these spectral features could be found elsewhere \cite{pal}. As in other 
transition metal oxide compounds, the major contribution to the physical 
properties of this system comes from the subtle changes in the states near the 
E$_F$ (within 2 eV from the E$_F$). A high resolution spectra of this region is 
shown in Fig. 3(a). Intensities of these spectra are normalized at regions above 
1.2 eV and below the E$_F$ and are shifted along the ordinate axis by a constant 
value for clarity.

Here, we concentrate on the subtle changes in the near E$_F$ spectral features, 
marked A and A' in Fig. 3. Different experiments \cite{saitoh,sarma,manas} and 
band structure calculations \cite{picket,ravindran,monodeep} have shown that both 
these features arise from the Mn $e_{g}$ spin-up states of the crystal field split 
MnO$_6$ octahedra. The spectra shown, taken at $77$ K, demonstrate the finer 
shifts in the spectral weight from A to A' as we go across the phase diagram shown 
in Fig. 1(a). The apparent gaps in the near E$_F$ DOS resulting from these shifts 
are usually called `pseudogaps'. As mentioned earlier, recent photoemission 
studies \cite{ebata1,pal,ebata2} have highlighted the importance of such 
pseudogaps, particularly for the models based on electronic phase separation 
\cite{moreo1,moreo2}. Fig. 3(a) shows that some finite number of states build up 
at A' as we go from FMI ($x = 0.3$) to AFMI ($x = 0.5$) composition. 
Further increase of $x$ to $0.7$ also results in an increased DOS at A'. Spectra 
from the $x$ = $0.85$ and $0.9$ samples display a distinct feature at A'. In panel 
(b) we have plotted the spectra corresponding to each composition together with 
that from the $x$ = $0.3$ for a comparison. It should be noted that, though all 
the samples with $x$ = $0.5$, $0.6$ and $0.7$ are AFMI at $77$ K the number of 
states at A' keep increasing with increase in $x$ showing that the pseudogap 
originating from these shifts exist also in the AFMI regime. This building up of 
DOS at A' could be associated with the growing FMM domains in the AFMI matrix in 
the Sm$_{1-x}$Ca$_x$MnO$_3$ system with progressive Ca doping. The increase in the 
DOS at A' should be due to the shift of states from A to A' following the 
growth of FMM domains. As we move on to $x$ = $0.85$ and $0.9$, the feature at A' 
becomes quite prominent. In these compositions, the FMM domains must be large 
enough leading to the metallic behavior shown by them. This has been confirmed 
earlier using neutron diffraction studies where we \cite{martin1} had shown that 
the Sm$_{1-x}$Ca$_x$MnO$_3$ system with $x$ = $0.85$ and $x$ = $0.9$ have a unique 
crystalline structure with FMM clusters embedded in a G-type AFMI background.

\begin{figure}[t]
\vskip 1.0cm
\includegraphics[width=3.0in]{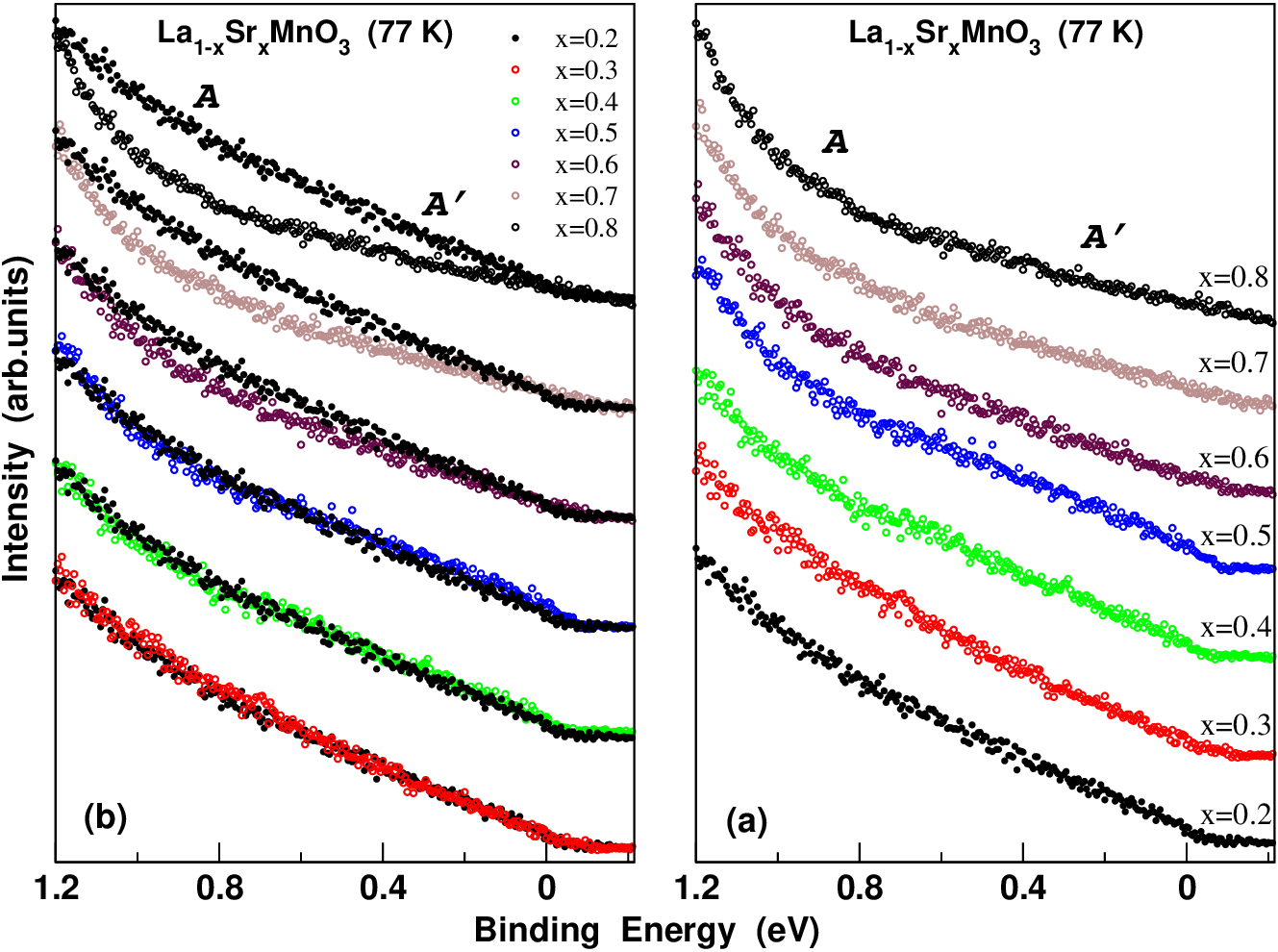}
\caption{\label{figure_4} Panel (a): High resolution photoemission spectra of the 
near E$_F$ features of La$_{1-x}$Sr$_x$MnO$_3$ compositions taken at $77$ K. (b): 
Spectra from different compositions plotted against that from $x$ = $0.2$ for 
comparison.}
\end{figure}

Let us now compare these results from Sm$_{1-x}$Ca$_x$MnO$_3$ with those from the 
canonical \cite{tokura} DE system, La$_{1-x}$Sr$_x$MnO$_3$. Figure 4 (a) shows the 
high resolution photoemission spectra of the near E$_F$ region of 
La$_{1-x}$Sr$_x$MnO$_3$ system taken at $77$ K. These spectra are also normalized 
and displayed in the same way as described earlier. The phase diagram 
\cite{chmaissem} of La$_{1-x}$Sr$_x$MnO$_3$ system (Fig. 1(b)) shows that 
compositions with $x$ = $0.2$ to $0.6$ are FMM while $x$ = $0.7$ and $0.8$ are 
AFMI. Correspondingly, our spectra from different metallic compositions also show 
the presence of a Fermi edge; the hallmark of metallicity. In Fig. 4 (b) we have 
plotted the spectra from different compositions together with that from the $x$ = 
$0.2$ sample. A close observation and comparison of this figure with Fig. 3 (b) 
will reveal that the spectra from different metallic compositions do not show any 
distinct feature corresponding to A' as in the case of the Sm$_{1-x}$Ca$_x$MnO$_3$ 
system. The MI transition in Sm$_{1-x}$Ca$_x$MnO$_3$ (from $x$ = $0.9$ to $x$ = 
$0.3$) was accompanied by the depletion of the feature at A' due to the transfer 
of some DOS from A' to A; namely the pseudogap. On the other hand, in case of the 
La$_{1-x}$Sr$_x$MnO$_3$ system we neither find any formation of such a feature at 
A' nor any shift of DOS. Here, the MI transition is manifested as the opening up 
of an insulator gap. The plot in Fig. 4 (b), showing the spectra from $x = 0.2$ 
(metallic) together with that from $x = 0.8$ (insulating) makes this point clear. 
One can see that the insulator to metal transition here is accompanied by an 
almost uniform increase in the DOS over the region between 1.2 eV and the E$_F$.

The differences shown by the near E$_F$ spectral changes associated with the 
doping dependent MI transitions in the Sm$_{1-x}$Ca$_x$MnO$_3$ and the 
La$_{1-x}$Sr$_x$MnO$_3$ systems could be relevant to the fundamental understanding 
of the mechanism of CMR. The insulator-metal transition in the 
La$_{1-x}$Sr$_x$MnO$_3$ system is due to the hopping of the e$_g$ electrons from 
Mn$^{3+}$ to Mn$^{4+}$ via the DE and superexchange interactions. These 
interactions are sensitive to the distortions in the MnO$_6$ octahedra which 
control the W and the electron correlation effects. With its small W these 
correlation effects are strong in case of the Sm$_{1-x}$Ca$_x$MnO$_3$. As 
mentioned before, the mixed phase compositions of this system have FMM domains in 
a CO-AMFI matrix \cite{martin1}. With its strongly distorted MnO$_6$ octahedra 
the C - type AFMI regions could be more prone to the electron localization effects 
compared to the FMM domains. The charge carriers (e$_g$ electrons) of the AFMI 
regions should thereby be more localized compared to the e$_g$ electrons in FMM 
regions. Consequently, this can result in higher binding energies for those e$_g$ 
electrons in AFMI regions compared to the ones in the FMM domains. As the sizes of 
these FMM domains increase with Ca doping, the number of itinerant e$_g$ electrons 
also go up resulting in a shift of spectral weight from A to A'. Conversely, this 
is the origin of the pseudogap. As Moreo et al. \cite{moreo1} have proposed 
earlier, this pseudogap could be a generic feature of such phase separated 
systems. But, the scenario is different in case of the La$_{1-x}$Sr$_x$MnO$_3$ 
system where such mixed phases do not exist. There the DE and superexchange 
processes through the undistorted Mn-O-Mn bond lead only to the hoping of e$_g$ 
electrons with no localization processes or pseudogaps involved.

\section{CONCLUSION}

In conclusion, we have shown one of the important differences in the near E$_F$ 
electronic behavior of phase separated and DE systems using high resolution 
photoemission experiments. Our study of the doping dependent MI transitions in 
Sm$_{1-x}$Ca$_x$MnO$_3$ compounds shows that the phase separated systems indeed 
show a pseudogap behavior in their near E$_F$ electronic spectrum over a large 
region of their phase diagram. A comparison of this system with the canonical DE 
La$_{1-x}$Sr$_x$MnO$_3$ system reveals the differences in the nature of the gaps 
associated with the MI transitions between these two systems. These differences 
could be ascribed to the distortions in the MnO$_6$ octahedra of their crystal 
structures which regulate the electron - electron and electron - lattice 
correlation effects determining the localization of charge carriers and thereby 
the pseudogap behavior.


\begin{thebibliography}{99}
\bibitem{zener}{C. Zener, Phys. Rev. {\bf 82}, 403 (1951).}
\bibitem{anderson}{P. W. Anderson and H. Hasegawa, Phys. Rev. {\bf 100}, 675 
(1955).}
\bibitem{gennes}{P. -G. de Gennes, Phys. Rev. {\bf 118}, 141 (1960).}
\bibitem{millis}{A. J. Millis, B. I. Shraiman and R. Mueller Phys. Rev. Lett. 
{\bf 77}, 175 (1996).}
\bibitem{tvr}{T. V. Ramakrishnan, H. R. Krishnamurthy, S. R. Hassan and 
G. Venketeswara Pai, page 417-441, Colossal Magnetoresistive Oxides, edited by T. 
Chatterji, Kluwer Academic Publishers, The Netherlands, 2004).}
\bibitem{moreo1}{A. Moreo, S. Yunoki, and E. Dagotto, Science {\bf 283}, 2034 
(1999).}
\bibitem{moreo2}{A. Moreo, S. Yunoki, and E. Dagotto, Phys. Rev. Lett. {\bf 83}, 
2773 (1999).}
\bibitem{simon}{Ch. Simon, S. Mercone, N. Guiblin, C. Martin, A. Brulet and G. 
Andre, Phys. Rev. Lett. {\bf 89}, 207202 (2002).}
\bibitem{martin1}{C. Martin, A. Maignan, M. Hervieu, B. Raveau, Z. Jirak, M. M. 
Savosta, A. Kurbakov, V. Trounov, G. Andre and F. Bouree, Phys. Rev. B {\bf 62}, 
6442 (2000).}
\bibitem{ling}{C. D. Ling, E. Gardano, J. J. Neumeier, J. W. Lynn and D. N. Argyriou, 
Phys. Rev. B {\bf 68}, 134439 (2003).}
\bibitem{radaelli}{P. G. Radaelli, R. M. Ibberson, D. N. Argyriou, H. Casalta, K. H. 
Andersen, S.-W. Cheong and J. F. Mitchell, Phys. Rev. B {\bf 63}, 172419 (2001).}
\bibitem{tokura}{Y. Tokura, Vol. 2, 2000. Colossal Magnetoresistive Oxides, edited 
by Y. Tokura, Advances in Condensed Matter Science (Gordon and Breach Science 
Publishers, Amsterdam, 2000).}
\bibitem{ebata1}{K. Ebata, H. Wadati, M. Takizawa, A. Fujimori, A. Chikamatsu, H.
Kumigashira, M. Oshima, Y. Tomioka, and Y. Tokura, Phys. Rev. B {\bf 74}, 64419,
(2006).}
\bibitem{pal}{ P. Pal, M. K. Dalai, R. Kundu, M. Chakraborty, B. R. Sekhar and C.
Martin, Phys. Rev. B {\bf 76}, 195120 (2007) and references therein.}
\bibitem{ebata2}{K. Ebata, M. Hashimoto, K. Tanaka, A. Fujimori, Y. Tomioka, and
Y. Tokura, Phys. Rev. B. {\bf 76}, 174418 (2007).}
\bibitem{jmm}{C. Martin, A. Maignan, M. Hervieu, B. Raveau, Z. Jirak, A. Kurbakov, V. Trounov, G. André, and F. Bourée, J. Magn. Magn. Mater. {\bf 205}, 184 (1999).}
\bibitem{martin2}{C. Martin, A. Maignan, M. Hervieu, C Autret, B. Raveau, and D. 
I. Khomskii, Phys. Rev. B {\bf 63}, 174402 (2001).}
\bibitem{chmaissem}{O. Chmaissem, B. Dabrowski, S. Kolesnik, J. Mais, J. D.
Jorgensen, and S. Short, Phys. Rev. B {\bf 67}, 94431 (2003).}
\bibitem{saitoh}{T. Saitoh, A. E. Bocquet, T. Mizokawa, H. Namatame, A. Fujimori, 
M. Abbate, Y. Takeda, and M. Takano, Phys. Rev. B {\bf 51}, 13942 (1995).}
\bibitem{sarma}{D. D. Sarma, N. Shanthi, S. R. Krishnakumar, T. Saitoh, 
T.Mizokawa, A. Sekiyama, K. Kobayashi, A. Fujimori, E. Weschke, R. Meier, G. 
Kaindl, Y. Takeda, and M. Takano, Phys. Rev. B {\bf 53}, 6873 (1996).}
\bibitem{manas}{M. K. Dalai, P. Pal, B. R. Sekhar, N. L. Saini, R. K. Singhal, K. 
B. Garg, B. Doyle, S. Nannarone, C. Martin, and F. Studer, Phys. Rev. B {\bf 74}, 
165119 (2006).}
\bibitem{picket}{W. E. Pickett, and D. J. Singh, Phys. Rev. B {\bf 53}, 1146 
(1996).}
\bibitem{ravindran}{P. Ravindran, A. Kjekshus, H. Fjellvag, A. Delin, and O. 
Eriksson, Phys. Rev. B. {\bf 65}, 064445 (2002).}
\bibitem{monodeep}{M. Chakraborty, P. Pal, and B. R. Sekhar, Solid State 
Commun. {\bf 145}, 197, (2008).}
\end{thebibliography}
\end{document}